
%
\input harvmac.tex
\overfullrule=0pt
\Title{LAVAL PHY-21-92, hepth@xxx/9203062}
{\vbox{
\centerline{ A Note on the KP hierarchy}}}
\vskip1cm
\centerline{Didier A.~Depireux}
\smallskip
\centerline{D\'epartement de Physique,}
\centerline{Universit\'e Laval,}
\centerline{Qu\'ebec, Canada G1K 7P4.}
\vskip .3in
{\bf Abstract:} Given the two boson representation of the conformal algebra
$\hat W_\infty$, the second Hamiltonian structure of the KP hierarchy, I
construct a bi-Hamiltonian hierarchy for the two associated currents. The KP
hierarchy appears as a composite of this new and simpler system.
The bi-Hamiltonian structure of the new hierarchy gives naturally all the
Hamiltonian structures of the KP system.
\vskip .3in

\Date{ Feb.1992.}

%
\def\frac#1#2{{#1 \over #2}}
\def\ha{{1 \over 2}}
\def\d{\delta}
\def\pa{\partial}
\def\j{\jmath\,}
\def\bj{\bar\jmath\,}

%

	\def\CH{{\cal H}}

\hyphenation{De-pi-reux}

\lref\JLG{J.-L. Gervais, Phys.Lett. \bf B160 \rm (1985) 277.}
\lref\Walg{A.B. Zamolodchikov, Theor. Math. Phys. \bf65 \rm(1985) 1205; A.B.
Zamolodchikov and V.A. Fateev, Nucl. Phys. \bf B280 \rm [FS18] (1987) 644;
V.A. Fateev and S.L. Lykyanov, Int. J. Mod. Phys. \bf A3 \rm(1988) 507.}
\lref\KdV{I.M. Gelfand and L.A. Dickey, Russ. Math. Surv. \bf30 \rm(1975) 77;
Funct. Anal. Appl. \bf10 \rm (1976) 259; B.A. Kupershmidt and G. Wilson,
Invent. Math. \bf62 \rm(1981) 403; V.G. Drinfel'd and V.V. Sokolov, J. Sov.
Math. \bf30  \rm (1985) 1975.}
\lref\Hamstruck{ P. Mathieu, Phys.Lett. \bf B208 \rm (1988) 101; I. Bakas,
Phys.Lett. \bf B213 \rm(1988) 313; \bf B219 \rm(1989)  (283); K. Yamagishi,
Phys.Lett \bf B205 \rm (1988) 466; P. Di Francesco, C. Itzykson and J.B.
Zuber, Comm. Math. Phys \bf 140 \rm (1991) 543.}
\lref\PRS{C.N. Pope, L.J. Romans and
X.Shen, Phys.Lett. \bf B236 \rm (1990) 173; Phys.Lett. \bf B242 \rm (1990)
401; Phys.Lett. \bf B245 \rm (1990) 72.}
\lref\YannisI{I. Bakas, Phys.Lett \bf B228 \rm (1989) 57.}
\lref\Sato{E. Date, M. Jimbo, M. Kashiwara and T. Miwa, ``\it Proc. of RIMS
Symposium on non-linear integrable systems\rm'', eds. M.Jimbo and T.Miwa,
(World Scientific, 1983); G.Segal and G.Wilson, Publ. IHES \bf 61 \rm (1985)
1.}
\lref\Wata{Y. Watanabe, Lett. Math. Phys. \bf 7 \rm (1983) 99; Ann. Mat.
Pura Appl. \bf 137 \rm (1984) 77.}
\lref\Dick{L.A. Dickey, Ann. N-Y Acad. Sc., 1987, 131.}
\lref\WYnine{F. Yu and Y.-S. Wu, UU-HEP-91/09; J.M. Figueroa-O'Farril, J. Mas
and E. Ramos, Phys. Lett. \bf B266 \rm (1991) 298.}
\lref\YannisII{I. Bakas and
E. Kiritsis, Nucl. Phys. \bf B343 \rm (1990) 185; Erratum \it ibid. \bf B350
\rm (1991) 512.}
\lref\YannisIII{I. Bakas and
E. Kiritsis, ``Beyond the Large-$N$ Limit: Non-Linear $W_\infty$
as Symmetry of the SL(2,R)/U(1) Coset Model'',
UCB-PTH-91/44, LBL-31213, UMD-PP-92-37.}
\lref\BKK{I. Bakas, B. Khesin and E. Kiritsis, UCB-PTH-91/48, LBL-31303,
UMD-PP-92-36.}
\lref\WYnineteen{F. Yu and Y.-S. Wu,  ``Nonlinear $\hat
W_\infty$ Current Algebra in the $SL(2,R)/U(1)$ Coset Model'',UU-HEP-91/19.}
\lref\BPRSXD{E.
Bergshoeff, C.N. Pope, L.J. Romans, E. Sezgin and X. Shen, Phys. Lett.  \bf
B245 \rm(1990) 447; D.A. Depireux, Phys. Lett. \bf B252 \rm(1990) 586.}
\lref\WYY{F. Yu and Y.-S. Wu, Phys.Lett. \bf B263 \rm (1991) 220; K.
Yamagishi, Phys.Lett. \bf B259 \rm (1991) 436.}
\lref\KO{See for instance B. Konopelchenko and W. Oevel, Loughborough
preprint, Mathematical Science Report A149.}
\lref\RS{See for instance A. Reiman and M. Semenov-Tyan-Shanskii, J. Sov.
Math. \bf 31 \rm(1985) 3399, and references therein.}
\lref\fractional{I. Bakas and D.A. Depireux, Int. J. Mod.
Phys. \bf A\rm (1992); Mod. Phys. Lett. \bf A6 \rm (1991) 1561; Erratum \it
ibid. \bf A6 \rm (1991) 2351.}
\lref\Pi{M. de Groot, T.J. Hollowood, J.L. Miramontes, IASSNS-HEP-91/19,
PUPT-1251; N. Burroughs, M. de Groot, T.J. Hollowood, J.L. Miramontes,
IASSNS-HEP-91/42, PUPT-1263; IASSNS-HEP-91/61, PUPT-1285. }
\lref\MSBDNS{L.J. Mason and G.A.J. Sparling, Phys. Lett. \bf A137 \rm (1989)
29; I. Bakas and D.A. Depireux, Mod. Phys. Lett. \bf A6 \rm(1991) 399; V.P.
Nair  and J. Schiff, CU-TP-521 (1991).}
\lref\narga{F. Narganes-Quijano, Int.J. Mod. Phys. \bf
A6 \rm (1991) 2611.}
\lref\Pierre{P. Mathieu and W. Oevel, Mod. Phys. Lett.
{\bf A6} (1991) 2397; D.A. Depireux and P. Mathieu, Int.J.Mod.Phys. \bf
A\rm(1992).}

\secno=-1
\newsec{Introduction, Conclusion and Open Problem}

There are deep relations between conformal field theory and non-linear
integrable systems. For instance, the classical Virasoro algebra forms the
second Hamiltonian structure of the usual KdV equation \JLG. More generally,
the extended conformal algebras $W_N$ \Walg\ were shown \Hamstruck\ to form
the second Hamiltonian structure of the generalized equations of KdV type
\KdV. Later, it was demonstrated \WYY\ that $W_{1+\infty}$ \PRS, a linear
deformation of the $W_N$ algebras in the large $N$ limit \YannisI, is
related to the first Hamiltonian structure of the KP hierarchy  \Wata. The
second Hamiltonian structure of the KP hierarchy \Dick\ was then shown
to be isomorphic to $\hat W_\infty$ \WYnine, a centerless deformation of
$W_\infty$.

In \YannisII, a two boson
representation of $W_\infty$ was given.  Yu and Wu then presented the
corresponding realization of $\hat W_\infty$, in terms of two bosons $\phi$
and $\bar\phi$ \WYnineteen\ (see also \YannisIII).
With this representation of the conformal algebra $\hat W_\infty$ (the
Hamiltonian structure for the integrable system formed by the KP hierarchy),
it is shown in this letter that it is possible to write down a hierarchy for
the currents $\j(x) = \pa_x \phi(x)$ and $\bj(x) = \pa_x \bar\phi(x)$. Given
this new hierarchy and the expression of the fields entering the KP hierarchy
in terms of $\j$ and $\bj$, it is shown that the flows of the KP hierarchy
themselves can be written as composites of the flows of the $\j,$ $\bj$
hierarchy.  This new hierarchy is proved to be bi-Hamiltonian and to possess
 an infinite number of local conserved quantities. The two simplest of the
possible Hamiltonian structures for the $\j,$ $\bj$ hierarchy generate
$W_{1+\infty}$ and $\hat W_\infty$ when expressed in terms of the KP fields;
the use of the recursion operator of the $\j,$ $\bj$ hierarchy allows one to
construct an infinite number of Hamiltonian structures for the KP
hierarchy. However, this construction shows that these new Hamiltonian
structures are all related.

Through examples, the existence of a zero-curvature or matrix formulation for
the new hierarchy is then conjectured. A complete zero-curvature formulation
would help understand some aspects of this new hierarchy, and in
particular, the results presented here give some hopes to obtain the KP
hierarchy as a reduction of the self-dual Yang-Mills equations in four
dimensions. Work on this is in progress.

\vskip1cm

\noindent{\it Some open problems:} The results presented in this letter and
the existence of a two fermion $\psi$ and $\bar\psi$  realization of
$W_{1+\infty}$\BPRSXD, the first Hamiltonian structure of the KP hierarchy,
suggest the existence of  a `` $\psi,$ $\bar\psi$ hierarchy'', the KP
hierarchy appearing as a composite of that $\psi,$ $\bar\psi$ hierarchy.
Also, two boson realizations of $W_N$ algebras exist \narga, and it should be
possible to translate the present work to the framework of the generalized KdV
equations, thus allowing one to rewrite the generalized KdV hierarchies \KdV\
and maybe even the fractional ones \fractional\Pi\Pierre\ as
composites of much simpler hierarchies. Preliminary calculations show that
such is indeed the case. Finally, this work might also help understand the
symmetries of the $c=1$, $d=2$ string.

 Let us now give
some notations. We will be using differential $\pa$ and  pseudo-differential
$\pa^{-1}$ operators, where $\pa^{-1}$ is an integration symbol defined by
$\pa \pa^{-1} a = \pa^{-1} \pa a = a.$ Where needed, we will use square
brackets to make clear how $\pa$ or $\pa^{-1}$ act. For instance, $\pa_x a(x)
= a(x) \pa_x  + a'(x)$ whereas $[\pa_x a(x)] = a'(x) $ (with the same
convention for $\pa_x^{-1}$). Also, to shorten the expressions of the Poisson
brackets, we write $\{a,b\}=c \d'$ for $\{a(x),b(x')\} = c(x) \pa_x \d(x,x').$
Finally, commas in indices usually denote differentiation so that $u_{i,t_r}$
means $[\pa_{t_r}~u_i].$

\newsec{Reformulating the KP hierarchy.}

\noindent{\it The KP hierarchy:}
The KP hierarchy is an integrable system consisting of an infinite number of
non-linear differential equations. It is usually formulated in terms of a Lax
operator $L$, spanned by an infinite number of fields, with an infinite tail
of integration or pseudo-differential symbols:
\eqn\la{
L = \pa + u_0 \pa^{-1}+ u_1 \pa^{-2} + \ldots = \pa + \sum_{i=0}^\infty u_i
\pa^{-i-1}\quad,\quad\pa = \pa/\pa_x~. }
The fields $u_i$ depend on $x$ as well as on an infinite number of time
coordinates $t_2,$ $t_3,$ $\ldots$ The $r^{th}$ member or $r^{th}$ flow
of the KP
hierarchy is given by
\eqn\lb{{\pa\over\pa t_r} L = [ (L^r)_+,L]\quad ,}
where as usual, $(L^r)_+$ denotes the differential part of $L^r.$ Each
value of $r\geq2$ gives rise to a non-trivial flow, in $(1+1)$ dimension
(i.e. for the variables $x$ and $t_r$). It is always possible to
think of $u_1,$ $u_2,$ $\ldots$ as auxiliary fields and to eliminate them by
simple integration to give rise to an equation in $u_0$ only.  The first few
flows for the lowest values of $r$ are
\eqnn\lc
$$
\eqalignno{
r=2:\quad u_{0,t_2} & = 2 u_{1,x} + u_{0,xx} \quad,\cr
u_{1,t_2} & = 2 u_{2,x} + u_{1,xx} + 2 u_0 u_{0,x}\quad, \cr
u_{2,t_2} & = 2 u_{3,x} + u_{2,xx} + 4 u_{0,x} u_1 - 2 u_0 u_{0,xx}
\quad.\qquad &\lc  \cr}$$
\eqnn\ld
$$
\eqalignno{
r=3:\quad u_{0,t_3} & = 3 u_{2,x} + 3 u_{1,xx} + u_{0,xxx} + 6 u_0 u_{0,x}
\quad, \cr
u_{1,t_3} & = 3 u_{3,x} + 3 u_{2,xx} + u_{1,xxx} + 6 (u_0 u_1)_x
\quad, \cr
u_{2,t_3} & = 3 u_{4,x} + 3 u_{3,xx} + u_{2,xxx} + 3 u_0 u_{2,x} + 9
u_{0,x} u_2 + \cr
& \qquad 6 u_1 u_{1,x} - 3 u_0 u_{1,xx} - 3 u_{0,xx} u_1 \quad.
& \ld \cr } $$ %

Eliminating $u_1$ and $u_2$ by \lc\ and plugging in the
first equation of \ld\ gives the usual KP {\it equation}:
$4u_{0,xt_3} =  (u_{0,xxx} + 12 u_0 u_{0,x})_x + 3 u_{0,t_2t_2}.$

{}From now on, we restrict ourselves to the KP {\it hierarchy} \lb.
Its flows commute with each other
and are Hamiltonian, i.e. they can always
be written as $u_{i,t_r} = \{u_i,\int\CH_r\}$ where $\{.,.\}$ is a definite
Poisson bracket and $\int\CH_r$ some Hamiltonian. There are two such
Hamiltonian structures: the first one was given by Watanabe \Wata. It was
recently shown \WYY\ to correspond to the conformal algebra $W_{1+\infty}$
with $c=0$. A second Hamiltonian structure was given by Dickey \Dick, which
was then related to $\hat W_\infty$ \WYnine, a centerless deformation of the
usual conformal algebra $W_\infty$. The Hamiltonian densities for the KP
hierarchy take a particularly simple form, namely
\eqn\lee{
\CH_r = {1\over r} \ res \ L^r~, }
where the residue $res$ of a pseudo-differential operator is the coefficient
of its $\pa^{-1}$ term. These are the same densities for either Hamiltonian
structure, that is we can rewrite \lb\ as $u_{i,t_r} =
\{u_i,\int\CH_{r+1}\}_1= \{u_i,\int\CH_r\}_2$, where $\{.,.\}_{1,2}$ denote
the first and second Hamiltonian structures respectively.

\vskip.5cm

\noindent{\it The representation of $\hat W_\infty$:}
In \WYnineteen, Yu and Wu constructed a free field realization of
$\hat W_\infty$ in terms of two scalar fields. To this end, one considers two
currents, $\j (x)$ and  $\bj (x)$, with Poisson brackets
\eqn\lf{
\{\j(x),\j(x')\}_2 = 0 ~,~ \{\j(x),\bj(x')\}_2 = \d'(x,x')
 ~,~ \{\bj(x),\bj(x')\}_2 = 0~.}
Introducing the generating functional
\eqn\lg
{L = \pa + \sum_{i=0}^{\infty} u_i \pa^{-i-1} = \pa + \bj {1 \over \pa - (\j +
\bj)} \j\quad,} %
one finds the expression of the $u_i$'s in terms of $\j$ and $\bj.$ The first
few explicit expressions are
\eqnn\lh
$$
\eqalignno{
u_0 & = \j\bj\quad,\quad u_1 = - \j'\bj + \j^2 \bj + \j \bj^2\quad, \cr
u_2 & = \j'' \bj - 3 \j \j' \bj - 2 \j' \bj^2 - \j \bj \bj' + \j^3 \bj + 2
\j^2 \bj^2 + \j \bj^3\quad,\cr
u_3 & = - \j''' \bj + 4 \j \j'' \bj + 3 \j'{}^2 \bj + 3 \j'' \bj^2 + 3 \j' \bj
\bj' + \j \bj \bj'' - 6 \j' \j^2 \bj - 9 \j \j' \bj^2 - \cr
& \qquad 3 \j^2 \bj \bj' - 3 \j' \bj^3 - 3 \j \bj^2 \bj' + \j^4 \bj + 3 \j^3
\bj^2 + 3 \j^2 \bj^3 + \j \bj^4\quad.   &\lh\cr}$$
Yu and Wu have shown that given \lf\ and \lh, one obtains a realization of
$\hat W_\infty$ in terms of $\j$ and $\bj,$
that is the brackets $\{u_i,u_j\}_2$ as calculated from \lf\ and \lh\ form the
$\hat W_\infty$-algebra.

\vskip.5cm

\noindent{\it A reformulation of KP:}
Spurred by these results, I tried to determine whether the KP hierarchy
itself could be written in terms of a hierarchy for the $\j$ and $\bj$ fields.
Let us first observe that \lg\ gives the simple formula
\eqn\li{
u_i = \bj [(-\pa + \j + \bj)^i \ \j]\quad.}
Using the expression \lee\ for the Hamiltonian, we can generate flows for the
$\j$ and $\bj$ fields, by using
\eqn\lj
{\vec\j_{,t_r} = \pmatrix{\j \cr \bj \cr} _{t_r} = P_2 \nabla_{\vec\j} \int
\CH_r\quad,}
where $\nabla_{\vec\j} = (\d/\d\j,\d/\d\bj)$ and $P_2$ is the Hamiltonian
structure corresponding to \lf,
\eqn\lk
{P_2 = \bordermatrix{&\j&\bj\cr \j&0&\pa\cr \bj&\pa&0\cr}\quad.}

The first few flows with their Hamiltonian densities are given by
\eqn\ltemp{r=1\quad:\CH_1 = \j\bj\quad,\quad\j_{,t_1} =
\j'\quad,\quad\bj_{,t_1}  = \bj'\qquad.}
\eqnn\ll$$
\eqalignno{r = 2\quad:\CH_2 & = - \j' \bj + \j^2 \bj + \j \bj^2 \quad,\cr
\j_{,t_2} & = ( - \j'  + \j^2  + 2 \j \bj)' \quad, \cr
\bj_{,t_2} & = (\phantom{-}  \bj' + \bj^2 + 2 \j \bj)' \quad.\qquad \qquad
\qquad & \ll\cr}$$ %
\eqnn\lm$$
\eqalignno{r = 3\quad:
\CH_3 & = \j'' \bj - 3 \j \j' \bj - 2 \j' \bj^2 - \j \bj \bj' +
\j^3 \bj + 3 \j^2 \bj^2 + \j \bj^3\quad,\cr
\j_{,t_3} & = (  \j'' - 3 \j \j' - 3 \j' \bj + \j^3 + 6 \j^2 \bj + 3 \j
\bj^2)'\quad,\cr
\bj_{,t_3} & = (  \bj'' + 3 \bj \bj' + 3 \j \bj'  + \bj^3 + 6 \j
\bj^2 + 3 \j^2 \bj )' \quad.& \lm\cr}$$
One can easily check that, for instance, the third flow of the KP hierarchy
given in \ld\ is given by computing the $t_3$ derivative of the $u_i$
fields, using the expressions \lh\ and the flows \lm. The same holds true for
the other flows. It would be  interesting to find a formula analogous to
\lee\ directly in terms of some scalar operator for the fields $\j$ and $\bj.$

\vskip.5cm

\noindent{\it The Hamiltonian structures:} the possibility to reformulate the
KP hierarchy in terms of $\j$ and $\bj$ is fairly clear, given the
two boson realization of $\hat W_\infty$. However, the existence
of a multi-Hamiltonian structure for that new hierarchy is much less obvious
and it is to that question that we now turn. We already know the Hamiltonian
structure $P_2$ given in \lk. Ascribing a weight of 1 to $\j,$ $\bj $ and
$\pa$, and writing down the most general self-adjoint operator of a given
weight,  one finds by brute force a unique Hamiltonian structure for each
weight:

\eqn\ln
{P_1 = \pmatrix{
\matrix{(\pa+\j-\bj)^{-1}\j + \cr
\quad \j(\pa+\bj-\j)^{-1}  \cr
\phantom{\ }}    &
\matrix{ -1 - (\pa+\j-\bj)^{-1}\bj - \cr
\quad \j(\pa+\bj-\j)^{-1}\cr
\phantom{\ }}       \cr
\matrix{ 1 - (\pa+\j-\bj)^{-1}\j - \cr
\quad \bj(\pa+\bj-\j)^{-1}}     &
\matrix{(\pa+\j-\bj)^{-1}\bj + \cr
\quad \phantom{\ha}\bj(\pa+\bj-\j)^{-1}}     \cr}\quad,}
\eqn\lo
{P_2 = \pmatrix{0&\pa\cr \pa&0\cr}\quad,}
\eqn\lp
{P_3 =
\pmatrix{\pa \j + \j \pa & - \pa^2 + \pa\j + \bj \pa \cr
\phantom{\ }             & \phantom{\ }              \cr
 \pa^2 + \j \pa +  \pa \bj\ &  \pa \bj + \bj \pa     \cr}\quad,}
\eqn\lq
{P_4 = \pmatrix{
\matrix{- \pa^2 \j + \j \pa^2 + 2 \j \pa \j + 2 \j \bj \pa + \cr
          2 \pa \j \bj + 2 \pa \j\pa^{-1} \j\pa\cr} &
\matrix{  \pa^3 - \pa^2 \j - \pa \j \pa - \bj \pa^2 - \pa \bj \pa +\cr
           \pa \j^2 + \bj^2 \pa + 2 \pa \j \bj + 2 \j \bj \pa  +\cr
           2 \pa \j \pa^{-1} \bj \pa \cr} \cr
\matrix{  \pa^3 + \j \pa^2 + \pa \j \pa + \pa^2 \bj + \pa \bj \pa +\cr
          \j^2 \pa + \pa \bj^2 + 2 \pa \j \bj + 2 \j \bj \pa  + \cr
          2 \pa \bj \pa^{-1} \j \pa \cr} &
\matrix{  \pa^2 \bj - \bj \pa^2 + 2 \bj \pa \bj + 2 \j \bj \pa + \cr
          2 \pa \j \bj + 2 \pa \bj\pa^{-1} \bj \pa \cr} \cr}~.}
These structures are such that $\vec\j_{,t_r} = P_i \nabla_{\vec\j} \int
\CH_{r+2-i},$ $i=1,2,3,4.$ As we will see shortly, there exists
an infinite number of such Hamiltonian structures, although only two of them
are independent. In spite of the non-locality of the Hamiltonian structures
$P_1$ and $P_4,$ the corresponding Hamiltonian  densities  are {\it local},
since they are polynomial in the $u_i$'s as seen from \lee, and hence
polynomial in the  fields $\j$ and $\bj$, as seen from \li.

$P_2$ and $P_3$ satisfy the Jacobi identities. We observe that $P_4 = P_3
P_2^{-1} P_3.$ Also, $P_3 = P_2 P_1^{-1} P_2$ (which, given the complicated
form of $P_1$ and $P_3$, is most easily checked in the form $P_2^{-1} P_1
P_2^{-1} P_3 = 1 $). Hence, just as in the KdV case \KO, the potential
Poisson brackets turn out to be related, and only two of them are independent.
The Jacobi identities are satisfied for all the $P_i$'s, since as we mentioned
they are satisfied by $P_2$ and $P_3.$
Note that there is no room for an independent third Hamiltonian structure,
since  the $P_i$'s given in \ln--\lq\ are uniquely defined.

Let us now see what Hamiltonian structures the $P_i$'s correspond to in terms
of the original KP fields $u_0$, $u_1$, $u_2$, $\ldots$
With respect to the first structure $P_1$,  we find
\eqnn\lr
$$\eqalignno{
\{u_0,u_0\}_1 & = 0 \quad,\quad \{u_1,u_0\}_1 = u_0 \d' \quad,\quad
\{u_1,u_1\}_1 = 2 u_1 \d' + u_1' \d \quad,\cr
\{u_0,u_2\}_1 & = u_0 \d'' + 2(u_1 + u_0')\d' + (2 u_1 + u_0')'\d\quad. & \lr
\cr} $$
With the field redefinitions
\eqn\lt
{W_1 = u_0\quad,\quad W_2 = u_1 + \ha u_0'\quad,\quad W_3 = u_2+u_1'+\
{\rm terms\ in\ } u_0 \quad,} %
these become
\eqnn\lu
$$\eqalignno{
\{W_1,W_1\}_1 & = 0 \quad,\quad \{W_2,W_1\}_1 = W_1 \d' \quad,\cr
\{W_2,W_2\}_1 & = 2 W_2 \d' + W_2' \d \quad,\quad \{W_1,W_3\}_1 = 2 W_2
\d' + 2 W_2' \d \quad,& \lu \cr }
$$
which are the first few brackets of $W_{1+\infty}$ with central charge $c=0$
\WYY. This was expected simply by looking at the dimensions, or from the fact
that $W_{1+\infty}$ has already  been shown to be the first Hamiltonian
structure of KP in its usual formulation. Note that $W_2$ is the stress
tensor. Alternatively, \ln\ can be thought of as a new realization of
$W_{1+\infty}$.

$P_2$ was already shown to give $\hat W_\infty$. In this case, the field
redefinitions that yield the $\hat W_\infty$ algebra, as given in \WYnine, are
\eqnn\lv
$$\eqalignno{
W_2 & = u_0\quad,\quad W_3 = u_1 + \ha u_0'\quad,\quad W_4 = u_2+u_1'+\ha
u_0^2+ \frac15 u_0''\quad,\cr
W_5 & = u_3 + \frac 32 u_2' + u_1'' + \frac14 u_0''' + 3 u_0 u_1 + \frac 32
u_0 u_0'\quad.\ & \lv \cr} $$
As for
$P_3$,  with the same field redefinitions as above (but with the index of the
$W$'s shifted as $W_i \rightarrow W_{i+1}$),
we find
\eqnn\lx
$$\eqalignno{
\{W_3,W_3\}_3 & = 4 W_4 \d' + 2 W_4' \d  \quad, \cr
\{W_3,W_4\}_3 & = W_3 \d''' + 2 W_3' \d'' + (5 W_5 + \frac 32 W_3'' + \frac 32
W_3^2 ) \d' + \cr
& \qquad (3 W_5 + \frac 25 W_3'' + \frac32 W_3^2)' \d \quad,\cr
\{W_3,W_5\}_3 & = \frac {14}5 W_4 \d''' + \frac {32}5 W_4' \d'' + (6 W_6 + 3
W_4'' - 6 W_3 W_4 ) \d' + \cr
& \qquad (4 W_6' - 4 W_3' W_4 - 2 W_3 W_4' ) \d \quad, \cr
\{W_4,W_4\}_3 & = 3 W_4 \d''' + \frac92 W_4' \d'' + (6 W_6 + \ha W_4'' - 6 W_3
W_4 ) \d' + \cr
& \qquad (3 W_6' - \ha W_4'' - 3 W_3 W_4 )' \d \quad.
& \lx \cr} $$
This is a new
Hamiltonian structure for the KP hierarchy.

In the case of the sl($N$) KdV hierarchies, we know that the so-called second
Hamiltonian structure corresponds to the classical $W_N$ algebra. The first
structure is then obtained by shifting the highest spin field by a constant,
that is under a shift of the highest spin field $u_N,$
$u_N \rightarrow u_N + \lambda,$ the second Hamiltonian structure gets
transformed as  $\{.,.\}_{2} \rightarrow \{.,.\}_{2}+ \lambda \{.,.\}_{1} .$
It would be interesting to know what the relation between
the Hamiltonian structures is, both for the KP and the $\j,$ $\bj$ hierarchy.
Of course we know that it is possible to find field redefinitions so that we
can go from $W_{1+\infty}$ to $W_\infty$ (see first work of \WYnine);
similarly, one can decouple the spin 2 field in $W_\infty$ to obtain an
algebra $W_{\infty-2}$ that contains fields of spins 3,4,$\ldots$ This
procedure can be repeated {\it ad infinitum} to derive an algebra with fields
of spins $N,$ $N+1,$ $\ldots$ The full structure of these algebras is given in
\BKK. Here $\hat W_\infty$ is a non-linear deformation of $W_\infty$, so it is
natural to expect that the algebra \lx\ is a deformation of $W_{\infty-2}$.
The first commutator of \lx\ supports this hypothesis.

It would therefore seem that the KP
hierarchy in its usual formulation \lb\ possesses an infinite number of
Hamiltonian structures $W_{1+\infty},$ $\hat W_\infty,$ $\hat W_{\infty-2},$
$\ldots$ When considered from the point of view of the $\j,$ $\bj$ formulation
however, only two of these structures turn out to be independent, all other
structures being generated by a recursion operator ($P_{i+1} P_i^{-1}$).
Also, it is not clear which of $P_1,$ $P_2$ or $P_2,$ $P_3$ should be taken as
the fundamental Hamiltonian structures. $P_1,$ $P_2$ correspond to the most
natural structures from the point of view of $W$-algebras, whereas the pair
$P_2,$  $P_3$ has the advantage of being local. Let us note however that this
locality of $P_2,$ $P_3$ is linked to the choice of our starting point, namely
the choice of $\j$ and $\bj$ as fundamental fields. In terms of a $\psi,$
$\bar\psi$ hierarchy of the type mentioned in the introduction, we expect the
Hamiltonian structures corresponding to $W_{1+\infty}$ and $\hat W_\infty$ to
be the local ones.

We see that we have been able to reformulate the KP hierarchy in terms of a
very simple one; the relation between the KP fields and the fields $\j$ and
$\bj$ takes the form \li, and the Hamiltonian structures of KP are summarized
by $P_2$ and $P_3.$

\newsec{Towards a zero-curvature formulation.}

The KP hierarchy is usually written in the form \lb, i.e.
in terms of a {\it scalar} pseudo-differential operator, but it is possible to
write it in terms of sl($\infty$) matrices \RS. In terms of $\j$ and $\bj$,
the matrix formulation should require only $2\times2$ matrices.

To find this matrix formulation, recall that in the formalism of Drinfeld
and Sokolov, the relation between the scalar Lax operator $L$ and the matrix
operator $\pa + Q$ is found for instance by writing $\Psi^T = (\psi_1,\psi_2,
\ldots,\psi_N).$ Then in the $N$ equations
\eqn\lz{ (\pa + Q) \Psi = 0 \quad,}
one eliminates $\psi_2, \ldots,\psi_N$ in terms of $\psi_1$. The resulting
equation is of the form $L \psi_1 = 0$, where $L$ is the scalar Lax operator
of the SL($N$) KdV hierarchy. Assuming the same holds true here, the most
natural matrix operator is
\eqn\lla{Q = \pmatrix{ 0 & \bj \cr -\j & -(\j + \bj)\cr}}
so that $(\pa + Q)\Psi = 0 $ is seen to reduce to $L \psi_1=0$, where $L$
corresponds to \lg. With the form \lla\ of $Q$ and imposing that the
zero-curvature condition
\eqn\llb{[\pa_x + Q, \pa_{t_r} + H_r ] = 0 \quad}
gives the $r^{th}$ flow of the $\j,$ $\bj$ hierarchy, one finds that the
Hamiltonians $H_r$  are
uniquely defined, and are explicitly given by
\eqn\llc{H_1 = \pmatrix{0 & \bj \cr -\j & -(\j + \bj)\cr}}
\eqn\lld{H_2 = \pmatrix{- \j \bj & \bj' + \j\bj + \bj^2 \cr
\j' - \j\bj - \j^2 & \j' - \bj' - 3 \j \bj - \j^2 - \bj^2 \cr}}
whereas for $H_3$, the matrix elements are given by
\eqnn\lle
$$
\eqalignno{(H_3)_{11} & = \j' \bj - \j \bj' - 2 \j^2 \bj -2\j \bj^2\quad, \cr
(H_3)_{12} & = \phantom{-} \bj'' + \j' \bj + 2 \j \bj' + 3 \bj \bj' + \j^2 \bj
+ 4 \j \bj^2 + \bj^3 \quad,\cr
(H_3)_{21} & = - \j'' + \j \bj' + 2 \j' \bj + 3 \j \j' - \j \bj^2 - 4 \j^2 \bj
- \j^3 \quad,\cr
(H_3)_{22} & = (H_3)_{21} - (H_3)_{12} + (H_3)_{11}\quad. & \lle \cr}$$
Let us stress that \llc--\lle\ were found by requiring \llb\ to reproduce the
flows \ll--\lm, for lack of a guiding principle. The existence and uniqueness
of $H$ shows it should be possible to derive the flows in a self-consistent
way from \llb\ {\it alone}. To that end, one would like to find an equivalent
to the expansion in a spectral parameter which is used in the zero-curvature
formulation of KdV flows; alternatively, one could start from the self-dual
Yang-Mills equations in four dimensions, reduced by two Killing symmetries (see
\fractional\MSBDNS\ for details). Starting in a space with metric $ds^2 = 2
dx\,dy + 2 dz\,dt,$ the  self-duality conditions on the field strength $F$ of
Yang-Mills theory, $F={}^*F$, become  \eqn\llf{
F_{xt} = 0 \quad,\quad F_{xy} = F_{zt} \quad,\quad F_{yz} = 0\quad.}
After reducing these equations with respect to $y$ and $z$, and making the
identification $A_x=Q,$ $A_t = H_r$, one finds, after imposing for instance
the equations of the second flow \ll, that $A_y$ and $A_z$ are not uniquely
defined,  but rather depend on 4 numerical factors.
This indicates that it should be
possible to obtain the KP hierarchy in the $\j,$ $\bj$ formulation as a
reduction of the self-dual Yang-Mills equations; the Ansatz to be imposed
on the connections is not clear yet. In all likelihood, the right Ansatz for
$Q$ is not the one given in \lla, but it must somehow be related to that form.

\vskip2truecm
{\bf Acknowledgements.}\par

I wish to thank I. Bakas with whom this work was initiated and P.
Mathieu for numerous discussions. I would also like to thank E. Kiritsis,
J. Schiff and Y.-S. Wu for discussions and reading the manuscript.

This work was supported by NSERC (Canada), FCAR
(Qu\'ebec), and BSR (Universit\'e Laval).

\listrefs
\end